\documentclass[journal,draftcls,onecolumn,12pt,twoside]{IEEEtranTCOM}

\normalsize
	\usepackage{pdfsync}
	\usepackage{preamble}
	
\begin{document}

\title{\smaller Deep Learning Framework for Hybrid Analog-Digital Signal Processing in mmWave Massive-MIMO Systems}

\author{Alireza Morsali,~\IEEEmembership{Student Member,~IEEE,}
  Afshin Haghighat~\IEEEmembership{Senior Member,~IEEE,}
	and Benoit~Champagne,~\IEEEmembership{Senior Member,~IEEE}
  \thanks{\smaller This work was supported by the Natural Sciences and Engineering Research Council of Canada (NSERC)
    and InterDigital Canada
    . }
\thanks{\smaller A.~Morsali and B.~Champagne are with the Department of Electrical and Computer Eng., McGill University, Montreal, Quebec, Canada
  ; A.~Haghighat is with InterDigital Canada
	(e-mail: alireza.morsali@mail.mcgill.ca; Afshin.Haghighat@InterDigital.com; benoit.champagne@mcgill.ca
  ).}}

\date{}
\maketitle
\vspace{-5em}

\begin{abstract}
  Hybrid analog-digital signal processing (\hp) is an enabling technology to harvest the potential of millimeter-wave (\mmw) massive-MIMO communications. In this paper, we present a general deep learning (DL) framework for efficient design and implementation of \hp-based \mm~systems. Exploiting the fact that any complex matrix can be written as a scaled sum of two matrices with unit-modulus entries, a novel  \emph{analog} deep neural network (ADNN) structure is first developed which can be implemented with common radio frequency (RF) components.
  This structure is then embedded into an extended hybrid analog-digital deep neural network (HDNN) architecture which facilitates the implementation of \mmw~\mm~systems while improving their performance. In particular, the proposed HDNN architecture enables \hp-based \mm~transceivers to approximate any desired transmitter and receiver mapping with arbitrary precision.
  To demonstrate the capabilities of the proposed DL framework, we present a new HDNN-based beamformer design that can achieve the same performance as fully-digital beamforming, with reduced number of RF chains.
  Finally, simulation results are presented confirming the superiority of the proposed HDNN design over existing hybrid beamforming schemes.
\end{abstract}

\begin{IEEEkeywords}
Hybrid beamforming, deep learning, deep neural networks, hybrid analog-digital beamforming, massive-MIMO, mmWave, beyond 5G (B5G), 6G.
\end{IEEEkeywords}

\section{INTRODUCTION}

\IEEEPARstart{M}{ultiple}-input-multiple-output (MIMO) technology has revolutionized modern wireless communications by unveiling its potential to increase transmission capacity through the deployment of multiple antennas at the transmitter and receiver sides of a communication link \cite{TelatarETT99}.
In recent years, asymptotic analysis has revealed that \emph{massive}-MIMO systems employing large scale antenna arrays, exhibit a linear increase in capacity with the minimum number of antennas employed at either the transmitter or receiver even in sparse scattering environments \cite{LarssonICM14,RusekISPM2013}. This property is of crucial importance for extending the applications of mmWave communications, which until recently had been only considered for indoor and fixed outdoor scenarios, in order to enable multi gigabit per second data rates \cite{SchumacherComMag2021,XingCOML2021,TorkildsonITWC11,DanielsTVT07}.

Indeed, mmWave signals experience severe path loss (due to atmospheric absorption) and high penetration loss compared with microwave signals, which has hindered their use in wireless cellular and local area networks. However, recent advances in mmWave hardware, combined with the capabilities of massive-MIMO and the availability of spectrum above 6 GHz have revived mmWave communications. Especially, the highly selective beam steering capabilities provided by large-scale antenna arrays and sophisticated beamforming\footnote{In practice, beamforming can be employed at both the transmitter and the receiver ends of a wireless link, where it is  referred to as precoding and combining, respectively.} algorithms can mitigate the intrinsic limitations of mmWave channels \cite{HurITC13}.

In the conventional fully-digital (FD) implementation of MIMO systems,
each antenna element is connected to a dedicated radio frequency (RF) chain. While this approach is suitable for commonly used small scale MIMO systems, it is not suitable
to mmWave massive-MIMO systems equipped with large number of antenna elements
due to the high production costs and power consumption of the associated RF circuitry. Therefore, although mmWave massive-MIMO is the prime technology for future generations of wireless networks (e.g., beyond 5G (B5G) and 6G), the implementation of such systems still faces many technical challenges, and to date remains a topic of ongoing research \cite{AlkhateebICM14,Lu18,LinJCOM19,Di_JSAC20}.

Hybrid analog-digital signal processing (\hp) is an ingenious and effective approach to facilitate the implementation of mmWave massive-MIMO transceivers \cite{ZhangITSP05,BogaleJWCOM16,SohrabiISTSP16,AyachITWC14,AlkhateebJWCOM16,YuJSTSP,Mai_WCNC_16,Nguyen17,Morsali_WCNC_19,LiICL16,JiangWCOM0,MorsaliCOML17,MorsaliGSIP19,MorsaliICASSP20,MorsaliACCESS20}. In an HSP transmitter, which includes hybrid beamforming (HBF) as a special case, a low-dimensional baseband signal (e.g., precoder output) is converted to RF and then mapped into a higher-dimensional signal for transmission by the antennas, where the mapping is achieved by an analog processing network comprised of basic RF components such as phase-shifters, combiners and dividers; in an HSP receiver, the dual operations are performed in reverse order.
Consequently, the \hp~transmitter/receiver structure requires a smaller number of RF chains for conversion between the digital baseband and analog RF domains, compared to its FD counterpart.
In this work, our aim is to develop and validate a novel deep learning (DL) framework for the efficient design and implementation of HSP-based \mmw~\mm~systems.

\vspace{-2em}
\subsection{Related Works}
\vspace{-1em}

One of the most prominent techniques for designing HSP~systems consists in minimizing the Euclidean distance between the desired FD processor and its hybrid counterpart, which is the objective function used for HBF design in \cite{AyachITWC14,AlkhateebJWCOM16,YuJSTSP,Mai_WCNC_16,Nguyen17,Morsali_WCNC_19,LiICL16,JiangWCOM0}. Particularly, in \cite{AyachITWC14,AlkhateebJWCOM16}, compressed sensing
techniques are employed to exploit sparse characteristics of the mmWave channels while in \cite{YuJSTSP,Mai_WCNC_16}, a manifold optimization algorithm and a simultaneous matrix diagonalization technique are introduced, respectively, to solve the design problem.
Channel sparsity  is also considered in \cite{SohrabiISTSP16},\cite{SohrabiJSAC17} where iterative orthogonalization algorithms are proposed for designing spectrally efficient HBF transceivers.
Gram–Schmidt orthogonalization is used in \cite{LiICL16} to design a robust hybrid combiner with low complexity for an uplink multi-user scenario.
The mean square error (MSE) is considered
as the performance metric in \cite{LinJCOM19}, where an alternating  minimization technique is used to design the HBF matrices.
Considering that closed-form expressions with fixed amount of calculation are often more attractive in applications, non-iterative design algorithms exploiting this type of solutions are proposed in \cite{KhalidCOML18,MoluTWC18}.
The authors in \cite{HO_JCAS_19} investigate the design and implementation (using CMOS process technology) of a low-complexity HBF based on orthogonal beamforming codebooks and a local search scheme.

Recently, in light of the huge success of machine learning and particularly deep learning (DL) in various fields of engineering, deep neural networks (DNNs) have attracted considerable attention among researchers for designing communication systems \cite{ZhuJSCA19,QinWCOM19,HuangWCOM20,AlkhateebAccess18, HuangTVT19,  LinWCOML20, SungACESS20}.
In \cite{AlkhateebAccess18}, a DL model is proposed for predicting the beamforming vectors at several distributed and coordinated base stations (BSs) by using received pilot signals. In particular, the signatures of the signals jointly received at the BSs with omni/quasi-omni directional beam patterns are used to learn and predict the RF beamforming vectors.
Another DL-based HBF approach for \mmw~\mm~is presented in \cite{HuangTVT19}, where an autoencoder is used to design the analog and digital precoders based on geometric mean decomposition.
In \cite{LinWCOML20}, the problem of maximizing spectral efficiency with hardware limitation and imperfect channel state information (CSI) is tackled by training a DNN to learn the optimum beamformers. Imperfect CSI is also considered in \cite{SungACESS20}, where multi-user DNN-based HBF design using codebooks is developed. Moreover, in order to apply this scheme to situations where the CSI is unknown, the concept of a reference RF beamformer is introduced.

Convolution neural networks (CNNs) have also been investigated for \hp~under various conditions \cite{ElbirCOML19,PekenTWCOM20,BaoACCESS20}. A CNN framework for the joint design of precoder and combiner is proposed in \cite{ElbirCOML19} where the network accepts channel matrices as input and produces analog and baseband beamformers as output.
In \cite{PekenTWCOM20}, three CNN architectures with different complexities are proposed to obtain approximations to the singular value decomposition (SVD), which are used in turn in the design of HBFs.
In \cite{BaoACCESS20}, a simplified hybrid precoding scheme is developed by considering the equivalent channel from the transmitter RF chains to the receiver RF chains. Based on this precoding approach, a novel CNN-based combiner architecture is proposed which can be trained to optimize the spectral efficiency under hardware limitations and imperfect CSI.

\subsection{Motivations and Contributions}

As mentioned above, the premise of \hp~is achieving the performance of FD systems with limited number of RF chains. However, this is not a trivial task mainly because of the constant modulus constraint on the analog beamformer entries, which causes the ensuing optimization problems to be non-convex.
Moreover, since in precursory HBF~works \cite{AyachITWC14,AlkhateebJWCOM16,YuJSTSP,Mai_WCNC_16,Nguyen17,Morsali_WCNC_19,LiICL16,JiangWCOM0}, the focus was placed on designing linear transformation matrices for the analog and digital beamformers, the recent DNN-based designs have followed the same structural guideline \cite{AlkhateebAccess18, HuangTVT19,  LinWCOML20, SungACESS20,PekenTWCOM20,ElbirCOML19,BaoACCESS20} which imposes fundamental limits on system performance. In particular, the minimum number of required RF chains must be greater or equal to the number of transmitted/received symbols or the rank of the FD matrix \cite{SohrabiISTSP16,BogaleJWCOM16}. However, some of these limitations could potentially be mitigated by considering non-linear transformations.

The so-called \emph{expressive power} of DNNs makes it possible to approximate continuous functions with arbitrary precision \cite{LUNIPS17}. In the context of HSP, this property can thus be exploited to obtain a non-linear transformation that ultimately requires a smaller number of RF chains compared to the linear case.
This motivates us to propose a general DL framework for efficient design and implementation of HSP-based massive-MIMO systems. To this end, we first develop a novel \emph{analog} deep neural network (ADNN) structure using conventional RF components as found in existing HBF systems, i.e., \cite{AyachITWC14,AlkhateebJWCOM16,YuJSTSP}. The ADNN is then embedded in a extended hybrid analog-digital deep neural network (HDNN) architecture with the goal of facilitating the implementation of mmWave massive-MIMO transceiver systems and improving their performance. The proposed HDNN architecture allows for accurate approximation of desired transmitter and receiver mappings in HSP-based massive-MIMO systems. While the proposed framework is quite general and can be used in different applications, our main focus in this work lies on the study of uplink and downlink beamforming in \mm~transmissions, for which we develop and investigate new HSP-based transceiver designs.
Specifically, our main contributions can be summarized as follows:

\begin{itemize}

	\item Considering the unit-modulus constraint of analog beamformers and inspired by the seminal works \cite{ZhangITSP05,BogaleJWCOM16,SohrabiISTSP16}, we present a technique for decomposing any given matrix with arbitrary complex entries into a scaled sum of two matrices with unit-modulus entries.

	\item This decomposition is exploited to conceive an ADNN structure comprised of common basic RF components, i.e.: phase-shifters, rectifiers, combiners/dividers and switches. This structure enables the efficient implementation of DNNs directly in the analog domain which to the best of our knowledge, has not been previously addressed.

  \item To reduce the number of required RF chains in HSP systems, a deep learning framework for \hp~design is then presented by introducing a novel HDNN architecture, comprised of a baseband digital DNN and an ADNN interconnected by means of RF chains.
	 \item To demonstrate the capabilities of the proposed framework, we present uplink and downlink beamformer designs for \mmw~\mm~systmes which achieve FD beamforming performance with limited number of RF chains. In particular, it is shown through these designs that the proposed HDNN framework is capable of reducing this number below the limits of conventional and existing DNN-based HBF designs.
	\item  Extensive simulation results are presented to demonstrate the advantages of the proposed HDNN-based beamfomer designs, which can achieve the same performance as their FD counterparts with reduced number of RF chains.
\end{itemize}

The paper is organized as follows. Section II, introduces the extended HSP-based massive-MIMO system model under consideration in this work. Section III develops the novel ADNN structure which is based on a representation theorem for complex matrices with unit-modulus entries. Our proposed HDNN framework for HSP system design is presented in Section IV while a supervised beamformer design based on this structure are developed in Section V. Simulation results are presented in Section VI and finally, Section VII concludes the paper.

\emph{Notations:} We use bold capital and lowercase letters to represent matrices and vectors, respectively. Superscripts $(.)^H$ and $(.)^T$ indicate Hermitian and transpose operations, respectively, while $\norm{\cdot}_F$ denotes the Frobenious norm of a matrix. ${\mathbf{I}}_n$ and $\mathbf{1}_n$ denote an identity matrix of size $n\times n$ and a vector of all ones with size $n\times 1$, receptively. The element on the $p^{th}$ row and $q^{th}$ column of matrix $\mathbf{A}$ is denoted by ${A}(p,q)$. % $\mathbf{A}=\text{diag}(a_1,a_2,\dots .,\ a_n$) represents a diagonal matrix, in which ${a_1,a_2,\dots .,\ a_n}$ are placed diagonally on the matrix $\mathbf{A}$.
A complex $n\times1$ Gaussian random vector $\mathbf{x}$ with mean vector $\mathbf{m}=\mathbb{E}\{\mathbf{x}\}$ and covariance matrix $\mathbf{R}=\mathbb{E}\{\mathbf{x}^H\mathbf{x}\}$ is denoted by $\mathscr{CN}(\mathbf{m},\mathbf{R})$. The greatest integer less than or equal to $x$ is denoted by $\lfloor x\rfloor$. The fields of real and complex numbers are denoted by $\mathbb{R}$ and $\mathbb{C}$, respectively. The magnitude and phase of a complex number $z\in \mathbb{C}$ are denoted by $|z|$ and $\angle z$, respectively. The function composition operation is denoted by $\circ$, i.e., $(f\circ g)(x)=f(g(x))$.

\vspace{-2em}
\section{System Model}
In this section, we first review the conventional linear HBF massive-MIMO structure and then introduce an extended system formulation which sets the ground for the development of our proposed DL framework for \hp~under more general conditions.

\begin{figure}[t]
	\centering
	\includegraphics[width = 0.8\linewidth]{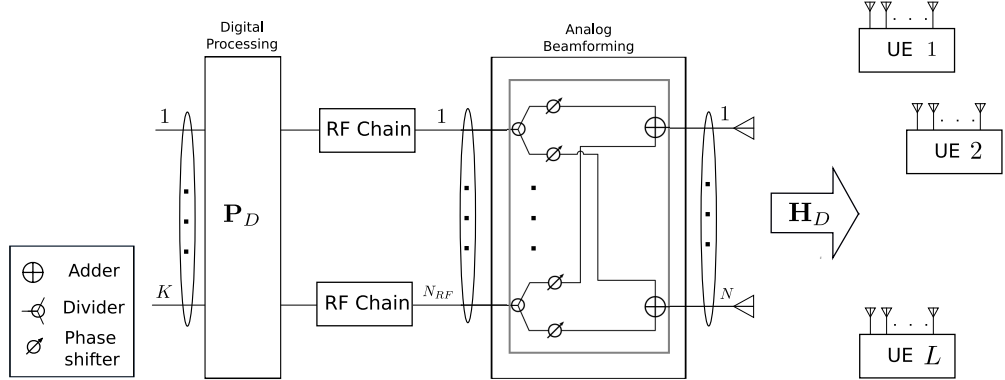}
	\caption{Massive-MIMO~transceiver with conventional HSP structure (downlink configuration).}
	\label{fig:HB}
  \vspace{-5em}
\end{figure}

\vspace{-2em}
\subsection{Conventional HBF Structure}\label{Sec:II_A}
Fig.~\ref{fig:HB} illustrates the typical downlink HBF configuration considered for the efficient realization of massive-MIMO BS transceivers in forthcoming wireless networks \cite{AyachITWC14,AlkhateebJWCOM16,YuJSTSP,Mai_WCNC_16,Nguyen17,Morsali_WCNC_19,LiICL16,JiangWCOM0}. The transceiver system consists of a baseband processor with $K$-dimensional input stream, a number $\nrf$ of RF chains, and an analog processor connected to $\nt$ antennas. Unless otherwise specified, we consider a multi-user scenario wherein the BS transceiver provides services to $L \le K$ user equipments (UE), each being allocated a fixed subset of the available data streams. For simplicity, we assume that the $l$th UE is equipped with $M_l$ antennas and the same number of RF chains, so that it is capable of FD processing, as our main focus lies on the design and realization of HSP-based massive-MIMO transceivers at the BS. In the sequel, we let $M=\sum_{l=1}^LM_l$ denote the total number of antenna elements employed by the UEs. While Fig.~\ref{fig:HB} focuses on dowlink transmission for simplicity, we herein consider both downlink and uplink connections, where the BS antennas are used for  transmitting to and receiving from the UEs, respectively. Below, we further expand the corresponding signal models.

\emph{Downlink:} The signal vector transmitted by the BS over one symbol duration $T_s$, denoted as $\xt\in\mathbb{C}^\nt$, can be expressed as
\begin{equation}\label{eq:hpconv}
\vspace{-1em}
\xt =  \sqrt{\rho}\,\pa\pd\s,
\end{equation}
where $\s=[s_{1},s_{2},...,s_{K}]^T\in \mathcal{A}^K$ is the input symbol vector with zero-mean random information symbols $s_k$ taken from a discrete constellation $\mathcal{A}\subset\mathbb{C}$ (e.g., M-QAM or M-PSK), normalized such that $\mathbb{E}\{\s\s^H\}=\I_\ns$, while matrices $\pd\in\mathbb{C}^{\nrft\times K}$ and $\pa \in\mathbb{U}^{ \nt \times \nrft}$ represent the digital and analog precoders, respectively, where
$\mathbb{U}=\{ z\in\mathbb{C}\colon|z|=1\}$. For normalization purposes, it is assumed that $\norm{\pa\pd}^2_F=1$, so that $\rho$ is the average transmit power.

The received signals at the UEs are represented by a concatenated signal vector $\xr\in\mathbb{C}^M$ which can be written as
\begin{equation}\label{murec1}
\vspace{-1em}
\xr=\Hd\xt+\nvec,
\end{equation}
where $\Hd\in\mathbb{C}^{ \nrue \times \nt}$ is a zero-mean random channel matrix representing flat fading transmission between the BS and the UE antennas, normalized such that  $\mathbb{E}\{\norm{\HH}^2_F\}=\nt\nrue$ without loss of generality, and
$\nvec\sim\mathscr{CN}(\mathbf{0},\mathbf{I}_\nrue)$ is an additive noise vector of size $\nrue$.
Each UE applies optimal FD linear processing on its received signal prior to the decoding stage. The linear processing performed by the multiple UEs is expressed as
\begin{equation}
\vspace{-1em}
\hat{\mathbf{s}}=\mathbf{C}\mathbf{y},
\label{decsymb}
\end{equation}
where $\mathbf{C}\in\mathbb{C}^{K\times\nrue}$ is a block diagonal combiner matrix.

\emph{Uplink:} To simplify the presentation, we use similar notations as in the donwlink case, with trivial modifications in vector dimensions as needed. Specifically, the received signal vector at the BS, denoted by $\xr\in\mathbb{C}^\nr$, is written as

\vspace{-2em}
\begin{equation}\label{murec2}
\xr=\Hu\xt+\nvec,
\vspace{-1em}
\end{equation}
where $\xt\in\mathbb{C}^M$ denotes the vector containing the concatenated transmitted signals from all UEs, $\Hu\in\mathbb{C}^{ \nt \times \nrue}$ is the zero-mean transmission matrix between the UEs and the BS antennas, normalized such that $\mathbb{E}\{\norm{\HH}^2_F\}=\nt\nrue$, and $\nvec\sim\mathscr{CN}(\mathbf{0},\mathbf{I}_\nr)$ is an additive noise term of size $\nr$.
In the conventional uplink HBF structure, a constrained form of linear processing is applied to the received signal, which is expressed as

\vspace{-2em}
\begin{equation}\label{eq:hpconv2}
\hat{\mathbf{s}} =  \ca\cd\xr,
\vspace{-1em}
\end{equation}
where matrices $\cd\in\mathbb{C}^{K\times \nrft}$ and $\ca \in\mathbb{U}^{ \nrft \times \nt}$ represent the digital and analog combiners, respectively. The final combiner output $\hat{\mathbf{s}}$ is then passed to a decoding stage, whose details fall outside the scope of this work.

\vspace{-2em}
\subsection{Generalized HSP Structure}
\label{Sec:GenFormul}

Here, we present an extended formulation for \hp\footnote{Although we focus on HSP at the BS to simplify presentation, similar aspects can be applied to the UE transceivers.} that generalizes the linear analog and digital transformations presented in Subsection~\ref{Sec:II_A}. This formulation provides an adequate basis for the subsequent development of our newly proposed DL-based HSP structures at both transmitter and receiver sides.

\emph{Downlink:} In a more general setting, the symbol vector $\s$ at the input of the HSP-
based massive-MIMO transmitter is first applied to a digital signal processor, whose output
is a baseband signal vector expressed as
\vspace{-2em}
\begin{equation}\label{fdhp}
\xbb=\fdt(\s)\in\mathbb{C}^\nrft,
\vspace{-1em}
\end{equation}
where $\fdt(\cdot)$ represents a generic mapping from $\mathcal{A}^K$ to $\mathbb{C}^\nrf$ or more concisely, $\fdt:\mathcal{A}^K\to\mathbb{C}^\nrf$. Then, $\nrf$ parallel RF chains convert the baseband signal vector $\xbb$ into a bandpass modulated RF signal vector $\xrf$. The latter is next input to the analog signal processing network whose output is the transmit signal vector, which is expressed as
\vspace{-1em}
\begin{equation}\label{fdhp1}
\xt=\sqrt{\rho}\,\fat(\xrf)\in\mathbb{C}^\nt,
\vspace{-1em}
\end{equation}
where $\fat:\mathcal{A}^\nrf\to\mathbb{C}^\nt$ is the corresponding mapping. The received signal at the UE is given by \eqref{murec1} and, in the case of FD linear combining, the demodulated symbols are obtained as in \eqref{decsymb}.

We emphasize that in the above formulation, the mappings $\fdt$ and $\fat$ are no longer limited to the class of linear transformations and could exhibit non-linear features, which is prerequisite for the consideration of DNNs.

\emph{Uplink:} The received signal from the large scale antenna array at the BS, which is given by \eqref{murec2}, first goes through the analog signal processing network, as represented by
\vspace{-1em}
\begin{equation}\label{fdhpr1}
\xrf=\far(\y)\in\mathbb{C}^\nrft,
\vspace{-1em}
\end{equation}
where $\far:\mathcal{A}^\nt\to\mathbb{C}^\nrf$ is the corresponding mapping. Then, $\nrf$ parallel RF chains convert the RF signal vector $\xrf$ into a digital signal vector $\xbb$ which is subsequently passed through a baseband processor, represented by
\vspace{-1em}
\begin{equation}\label{fdhpr}
\hat{\mathbf{s}}=\fdr(\xbb)\in\mathbb{C}^\ns,
\vspace{-1em}
\end{equation}
where $\fdr:\mathcal{A}^\nrf\to\mathbb{C}^K$ is the corresponding mapping.
Here again, we envisage a general configuration where the mappings $\far$ and $\fdr$ could be non-linear.

\vspace{-1em}
\section{Analog (RF) Deep Neural Networks} \label{Sec:SU}
In this section, we present a novel ADNN structure for efficient implementation of DNNs directly in the analog domain. We start by introducing terminology for DNN architecture and training that will serve as basis in subsequent developments. We then present the linear and non-linear RF modules which are required to carry out the necessary operations in DNNs. Finally we present the proposed ADNN structure.

\vspace{-1em}
\subsection{Deep Neural Networks}
\label{Sec:IIIDNN}
We consider complex-valued DNNs since in communication systems, complex numbers or vectors are used for the representation of the transmitted and received bandpass signals.
The mathematical expression of a complex-valued multi-layer perceptron (MLP), or artificial neural network, with one hidden layer, $d$ neurons and a linear layer at its output, is given by:
\vspace{-1em}
\begin{equation}
  \label{eq:per1}
  \ann(\x;\tet)=\W\,\psi_d(\afn(\x)),
\vspace{-1em}
\end{equation}
where $\x\in\mathbb{C}^{n}$ is the input vector, $\afn(\x): \mathbb{C}^n\to \mathbb{C}^d$ is an affine transformation, $\psi_d$ is a non-linear activation function and $\W\in\mathbb{C}^{m\times d}$ is the weight matrix of the output layer. Specifically, $\afn(\x)=\A\x+\b$ where $\A\in\mathbb{C}^{d\times n}$, $\b\in\mathbb{C}^{d}$ are the weight matrix and bias vector of the hidden layer, respectively. Moreover, for a given scalar activation function $\psi: \mathbb{C} \to \mathbb{C}$, the element-wise activation function $\psi_d: \mathbb{C}^{d}\to \mathbb{C}^{d}$ is defined as
  $\psi_d(\boldsymbol{\xi})=[\psi(\xi_{1}),\psi(\xi_{2}),\dots,\psi(\xi_{d})]^T$.
where $\boldsymbol{\xi}\in\mathbb{C}^{d}$.
Consequently, the set of parameters characterizing the MLP can be written as $\mathbf{\tet}=\{\W,\A,\b\}$.
A common practice for simplifying the implementation of DNN algorithms is to remove the bias term by adding an additional unit element to $\x$ and a column to $\A$. Consequently, by defining $\Aa=[\A, \b]$ and $\xa=[\x, \,1]^T,$ we can rewrite \eqref{eq:per1} as:
\vspace{-1em}
\begin{equation}
  \label{eq:mlp}
  \ann(\x;\tet)=\W\,\psi_d (\Aa\xa).
\vspace{-1em}
\end{equation}

To simplify the presentation, we focus on feed-forward DNN constructed from MLP with multiple hidden layers, but extensions to other types of networks are possible.
Let us consider a DNN comprised of $L\in\mathbb{N}$ hidden layers indexed by $l \in \{1,2,\ldots,L\},$ $n_l\in\mathbb{N}$ neurons at the $l$-th hidden layer, and one output layer indexed by $l=L+1$, with input vector $\x \in \mathbb{C}^{N_0}$ and output vector $\y \in \mathbb{C}^{n_{L+1}}$. The mathematical representation of such DNN can be written as

\vspace{-1em}
\begin{equation}
  \label{eq:dnn}
\y= \dnn(\x;\mathbf{\tet})=(\afn_{L+1}\circ\psi_{n_L}\circ\afn_L\circ\dots\circ\psi_{n_1}\circ\afn_1)(\x)
\vspace{-1em}
\end{equation}
where $\afn_l: \mathbb{C}^{n_{l-1}}\to \mathbb{C}^{n_l}$ is the affine transformation used at the $l$-th layer, i.e., $\afn_l(\x)=\A_l\x+\b_l$, with $\A_l\in\mathbb{C}^{n_l\times n_{l-1}}$, $\b_l\in\mathbb{C}^{n_l}$ representing the weights and biases of the $l$-th layer.
The ordered set of all parameters characterizing the DNN in \eqref{eq:dnn} can be written as $\mathbf{\tet}= \{\A_1,\b_1,\A_2,\b_2,\dots, \A_{L+1},\b_{L+1}\}$.

In practice, the network parameters in $\mathbf{\tet}$ are tuned by training the DNN on a given dataset of size $N_D$, represented here by $\pazocal{D}=\{(\x^{(i)},\y^{(i)})\}_{i=1}^{N_D}$ where, $\y^{(i)}$ is the desired output for given input vector $\x^{(i)}$. In batch learning, the data set is partitioned into $N_B$ mini-batches $\pazocal{B}_j$ such that $\pazocal{D}=\cup_{i=j}^{N_B}\pazocal{B}_j$ where each mini-batch is comprised of $N_D/N_B$ distinct samples drawn from the dataset.
The network parameters are randomly initialized and subsequently adjusted by means of an iterative process in two-steps: 1) the output of the network is first calculated for each mini-batch (forward propagation), and a loss value is obtained using a cost function based on the error between the actual outputs of the DNN and the desired outputs; 2) the parameters are updated based on the gradient of the cost function with respect to each element of $\tet$ (backward propagation).
An epoch consists of updating the network parameter $\tet$ through forward and backpropagation for all $N_B$ mini-batches. The training process is stopped after a number of epochs which is determined based on a pre-selected accuracy measure.

\vspace{-1em}
\subsection{Linear RF Modules}
\label{Sec:IIIA}
As discussed above, two different types of operations are required for calculating the output of each DNN layer, i.e., for the $l$-th layer, a linear transformation $\afn_l$ and an elementwise non-linear function $\psi_{n_l}$.  Let us first focus on the implementation of the linear transformation, which is not an immediate task when using phase-shifters and power-combiners/dividers in RF domain. In fact, most of the literature on hybrid beamforming is dedicated to this issue, and especially the non-convexity of design problems involving transformation matrices with unit-modulus entries.
Motivated by hybrid designs in \cite{ZhangITSP05,BogaleJWCOM16,SohrabiISTSP16}, we present below a first proposition which paves the way for developing an analog structure consisting of phase-shifters and power-combiners/dividers that can perform an arbitrary linear transformation. \begin{theorem}\label{Theo:asli}
	Any given complex matrix $\mathbf{A}\in\mathbb{C}^{l\times n}$ can be written as a scaled sum of two matrices $\mathbf{R}^{(1)},\mathbf{R}^{(2)}\in\mathbb{U}^{l\times m}$, with unit-modulus entries, i.e.,
	\begin{equation}
	\mathbf{A}=c(\mathbf{R}^{(1)}+\mathbf{R}^{(2)}),
	\end{equation}
	for some positive constant  $c\geq\frac{1}{2}\lVert \text{vec}(\mathbf{A})\lVert_\infty$.
\end{theorem}
\begin{IEEEproof}
	It is sufficient to show that for all $p=1,2,...,l$ and $q=1,2,...,n$, we have
	\begin{equation}
	A(p,q) =c\big(R^{(1)}(p,q)+R^{(2)}(p,q)\big).
	\end{equation}
	Since $\mathbf{R}^{(1)},\mathbf{R}^{(2)}\in\mathbb{U}^{N_R\times \ns}$, the above equation can be further written as
	\begin{equation}\label{eq:prof1}
	A(p,q) =c\big(e^{j\theta^{(1)}_{p,q}}+e^{j\theta^{(2)}_{p,q}}\big),
	\end{equation}
	where $e^{j\theta^{(1)}_{p,q}}$ and $e^{j\theta^{(2)}_{p,q}}$ represent the entries on the $p^{th}$ row and $q^{th}$ column of $\mathbf{R}^{(1)}$ and $\mathbf{R}^{(2)}$, respectively.
	Since $2c$ is greater than the absolute value of all the entries of $\mathbf{A}$, from Theorem 1 in \cite{ZhangITSP05}, there exist non-unique $\theta^{(1)}_{p,q}$ and $\theta^{(2)}_{p,q}$ such that \eqref{eq:prof1} holds which proves the theorem.
\end{IEEEproof}

Note that for a given matrix $\A$, the matrices $\da^{(1)}$ and $\da^{(2)}$ such that $\A=c(\da^{(1)}+\da^{(2)})$ holds are non-unique; below, we present a simple way to construct $\da^{(1)}$, $\da^{(2)}$.
By expressing the elements of $\A$ in polar form as $A(p,q)=|A_{p,q}|\exp({j\vartheta_{p,q}})$, we first compute
\begin{equation}\label{eq:digit}
c=\frac{1}{2}\max_{p,q}|A_{p,q}|.
\end{equation}
We then calculate the entries of $\da^{(1)}$ and $\da^{(2)}$ as
\begin{subequations}\label{eq:onesol}
	\begin{align}
		R^{(1)}(p,q)&=e^{j \big(\vartheta_{p,q}  + \cos^{-1} (\frac{|A_{p,q}|}{2c})\big) } \\
		R^{(2)}(p,q)&=e^{j \big(\vartheta_{p,q}  - \cos^{-1} (\frac{|A_{p,q}|}{2c})\big) }.
	\end{align}
\end{subequations}
By simple mathematical manipulations one can easily verify the validity of the presented solutions in \eqref{eq:digit} and \eqref{eq:onesol} (see also \cite{SohrabiISTSP16}).

\vspace{-1em}
\subsection{Nonlinear RF Modules}
Next, we discuss possible structures for the realization of non-linear activation functions in the analog domain with common RF modules.

 In \cite{hornik_91}, it is shown that an arbitrarily wide MLP with one hidden layer is a universal approximator due to the presence of the non-linear activation function. DNNs inherit this property and different conditions for the activation functions have been presented under which the universality of DNNs is satisfied \cite{hornik_91,park_91,lensho_93,huang_06,Mhaskar16}. In \cite{hornik_91}, the required conditions for the activation functions $\psi: \mathbb{R}\to \mathbb{R}$ are given as being continuous, bounded and non-constant, while these conditions are simplified to non-polynomiality in \cite{lensho_93}. Universality of DNNs is also studied in \cite{hanin2019,hanin2017} for the rectified linear unit (ReLU) activation function \cite{Nair10}, i.e.,
 \begin{figure}[t]
   \centering
 \includegraphics[width=\linewidth]{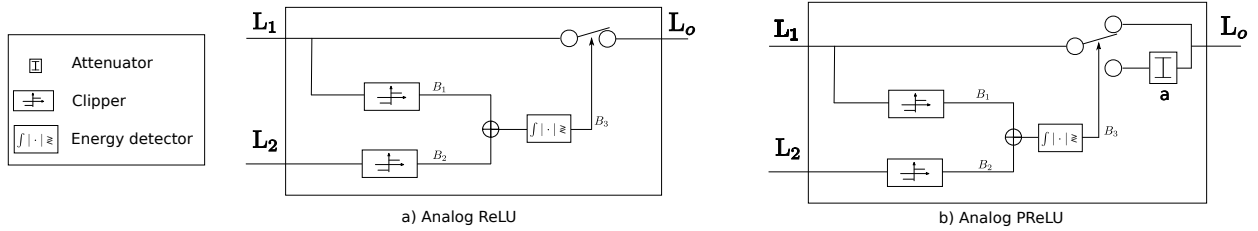}
 \caption{Block diagram of analog ReLU and PReLU activation functions. The input is applied to $L_1$, the carrier is applied to $L_2$, and the output is obtained from $L_o$.}
 \label{fig:AReLU}
 \vspace{-3em}
 \end{figure}

\vspace{-1em}
\begin{equation}
  \label{eq:relu}
\text{ReLU}(x)=x^+=\max (0,x),\quad x \in \mathbb{R}.
\vspace{-1em}
\end{equation}
Since the introduction of ReLU, several variations have been proposed among witch the leacky-ReLU and parametric ReLU (PReLU) have attracted considerable attention \cite{Maas13,He2015}.
In PReLU the vanishing part of the response is replaced by a nonzero linear section, i.e.,
\vspace{-0.5em}
\begin{equation}
  \label{eq:prelu}
\text{PReLU}(x)=\begin{cases} ax &\mbox{if } x<0 \\
x & \mbox{if } x \geq 0 \end{cases}
\vspace{-0.5em}
\end{equation}
where the linear slope $a>0$ can be learned along with other DNN parameters.
In complex-valued DNNs, the activation function must also be a complex-valued function. Based on  \cite{Trabelsi2018DeepCN}, the definition of \crel~can be extended to complex numbers as
\vspace{-1em}
\begin{equation}
  \label{eq:crel}
 \text{ReLU}(z)=\text{ReLU}(\mathfrak{R}(z))+j\text{ReLU}(\mathfrak{I}(z)),\quad z\in\mathbb{C},
\vspace{-1em}
\end{equation}
with similar extension for PReLU.

Since the baseband signal $\xbb$ is modulated on a sinusoidal carrier, both the in-phase (I) and quadrature (Q) components of the RF signal oscillate between a negative and positive peak value, regardless of the sign of $\xbb$. Consequently, unlike the baseband case, ReLU cannot be directly implemented with a single rectifier (e.g., diode). Herein, we therefore present a novel analog ReLU module which can realize \eqref{eq:crel} in the RF domain using basic circuit components.

The conceptual block diagram of the proposed analog ReLU module is shown in Fig.~\ref{fig:AReLU}(a). For complex-valued signals, this module is in effect applied to both the I and Q components. The modulated signal is fed to $L_1$ and the carrier is fed to $L_2$ as a reference. The output $L_o$ is equal to the input signal $L_1$ when the baseband signal is positive and zero otherwise.
Indeed, when the modulated signal $L_1$ and the carrier $L_2$ are in phase, corresponding to a positive value of the baseband component, $B_1$ and $B_2$ are equal and the energy of their sum is non-zero. Consequently, the output of the energy detector $B_3=1$ which can be used to activate (i.e., close) the switch. When the modulated signal $L_1$ and the carrier $L_2$ are out of phase (corresponding to a negative value of the baseband component), we have $B_2=-B_1$. Consequently, the energy of their sum is zero and the output of energy detector $B_3=0$, which opens the switch. Similarly, an analog PReLU module can be designed as shown in Fig.~\ref{fig:AReLU}(b) where the attenuator is a passive component which is designed based on the learned value of $a$ in \eqref{eq:prelu}.

\vspace{-1em}

\subsection{Analog Deep Neural Network (ADNN)}
\label{Sec:IIIB}
The RF modules conceived in the previous two subsections can now be used to achieve our main objective, which is to design an analog DNN structure, i.e. ADNN, such that
\begin{equation}
  \label{eq:objadnn}
  \begin{split}
  \fat(\xrf)&=\dnn_T(\xrf;\tet^A_T)\\
  \far(\xr)&=\dnn_R(\xr;\tet^A_R),
  \end{split}
\end{equation}
where $\dnn_T()$ and $\dnn_R()$ represent DNNs with respective parameter set $\tet_T^A$ and $\tet_R^A$.
As discussed in Subsection \ref{Sec:IIIDNN}, DNNs are formed by concatenating several MLPs. Consequently, we first focus on the RF implementation of an MLP with single hidden layer in \eqref{eq:mlp}. For developing the ADNN, we consider commonly used RF components in existing \hp, i.e., phase-shifters, and power-dividers (which also work as combiners), along with the simple analog ReLU or PReLuU modules introduced previously.
\begin{figure}[t]
  \centering
\includegraphics[width=0.7\linewidth]{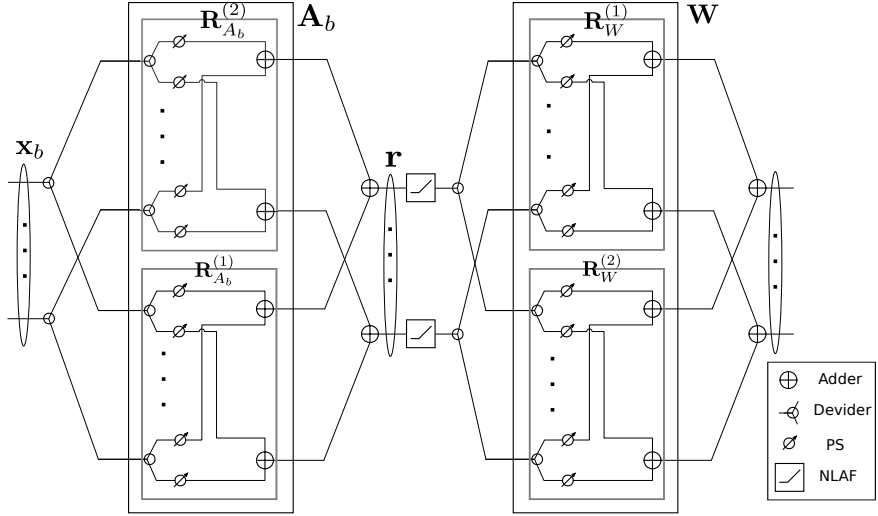}
\caption{Analog MLP (NLAF = non-linear activation function, i.e. ReLU or PReLU).}
\label{fig:ANN}
\vspace{-3em}
\end{figure}

\begin{theorem}\label{Theo:realization}
	A given MLP $\ann(\x;\tet)$ with one hidden layer and non-linear activation function $\psi_d()$ (either ReLU or PReLU) can be realized by the RF structure shown in Fig.~\ref{fig:ANN}.
\end{theorem}
\begin{IEEEproof}
  As discussed earlier an MPL is defined by matrices $\Aa$ and $\W$ which are used to calculate $\ann(\x;\tet)=\W\,\psi_d(\Aa\xa)$ where here, $\psi_d(\x)=\text{ReLU}(\x)$ or $\text{PReLU}(\x)$. Using Theorem \ref{Theo:asli}, there exist matrices $\mathbf{R}_{\Aa}^{(1)}$, $\mathbf{R}_{\Aa}^{(2)}$, $\mathbf{R}_{\W}^{(1)}$ and $\mathbf{R}_{\W}^{(2)}$ with constant-modulus entries, and positive scalars $c_A$, $c_W$, such that
\vspace{-1em}
  \begin{subequations}
    \vspace{-1em}
    \begin{align}
      \Aa=c_A(\mathbf{R}_{\Aa}^{(1)}+\mathbf{R}_{\Aa}^{(2)})\label{eq:Aa}\\ \W=c_W(\mathbf{R}_{\W}^{(1)}+\mathbf{R}_{\W}^{(2)})\label{eq:W}.
\vspace{-1em}
    \end{align}
  \end{subequations}
  As can be seen from Fig.~\ref{fig:ANN}, $\xa$ fist goes through dividers which feed the analog subnetowrks $\mathbf{R}_{\Aa}^{(1)}$, $\mathbf{R}_{\Aa}^{(2)}$, whose output are finally combined to produce vector $\r$. Since we operate in the analog domain, the power of the signal is split between each branch, so that only  $\frac{\xa}{\sqrt{2}}$
  is effectively applied on each branch; similar, a factor of $\frac{1}{\sqrt{2}}$ must be include in the ouput of the combiner to account for properties of the corresponding S-matrix \cite{pozar_2005}. Consequently, the vector of input signals to the ReLUs can be written as
\vspace{-1em}
  \begin{equation}
    \label{eq:powad}
  \mathbf{r} = \frac{1}{\sqrt{2}}(\mathbf{R}_{\Aa}^{(1)}\frac{\xa}{\sqrt{2}}+\mathbf{R}_{\Aa}^{(2)}\frac{\xa}{\sqrt{2}}).
\vspace{-1em}
  \end{equation}

  By substituting \eqref{eq:Aa} in \eqref{eq:powad}, we arrive at
\vspace{-1em}
  \begin{equation}
    \label{eq:Z}
  \mathbf{r} = \frac{1}{c_A}\Aa\xa.
\vspace{-1em}
  \end{equation}
  After passing through the ReLU and performing similar steps as above steps for $\W$, the output of the analog MLP can be written as
  \begin{equation}
    \label{eq:prof12}
  \ann(\x;\tet)=\frac{1}{c_Ac_W}\W\,\psi_d(\Aa\xa)=\W\,\psi_d(\Aa\frac{1}{c_Ac_W}\xa),
  \end{equation}
  where the second equality is due to the positive homogeneity of $\text{ReLU}$ i.e., $c\,\text{ReLU}(\x)=\text{ReLU}(c\x)$ for any positive $c$ ($c\in\mathbb{R}^+$).
\end{IEEEproof}

\begin{corollary}
Any given DNN $\dnn(\x;\mathbf{\tet})$ \eqref{eq:dnn} can be realized in the RF domain by concatenating several analog MLPs, as presented in Theorem \ref{Theo:realization}.
\end{corollary}

\section{Deep Learning Framework for HSP}
\label{Sec:DL}
In this section, by taking advantage of the proposed ADNN structure, we present a general DL framework for \hp~which enables hybrid systems to exhibit similar performance as FD systems with limited number of RF chains. These results are in contrast with conventional hybrid systems where, due to non-convex constraints imposed on the RF beamfomer, degraded performance must be accepted for reducing the number of RF chains. Moreover, our approach enables further improvements over existing DNN-based \hp~designs by incorporating ADNNs. In the existing DNN-based hybrid designs, deep learning is used for designing the analog and digital beamformers which face the same limitations as the conventional hybrid design, where the minimum number of required RF chains is equal to the number of symbols streams.
The proposed HDNN, however, is not bound to this constraint and as will be shown for selected cases, the number of RF chains can be further reduced significantly while maintaining the same performance as the FD systems.

\subsection{Proposed DL~Framework for \hp}
\emph{Downlink:}
Let us consider the downlink signal model introduced in Section ~\ref{Sec:GenFormul} for the generalized HSP structure. In the proposed HDNN framework, the non-linear digital processor in \eqref{fdhp} indeed corresponds to a first DNN, whose output is the baseband signal vector $\xbb$ i.e.,
\begin{equation}
  \label{eq:fatann}
  \fdt(\s)=\dnn_T^D(\s;\tet^{D}_T),
\end{equation}
where $\tet_T^{D}$ is the parameter set of this \emph{digital} DNN.
The baseband output of this DNN is then converted to the RF domain where it is represented by $\xrf$.
Subsequently, the non-linear operator in \eqref{fdhp1} is realized by means of a second ADNN structure, whose output (up to a scale factor) is the desired transmit signal vector $\x$, i.e.,
\begin{equation}
  \label{eq:fatann}
  \fat(\xrf)=\dnn_T^A(\xrf;\tet_T^A),
\end{equation}
where $\tet_T^{A}$ is the parameter set of the ADNN. % \deleted{ and $\zrf$ is the input of
\begin{figure}[t]
  \centering
\includegraphics[width=\linewidth]{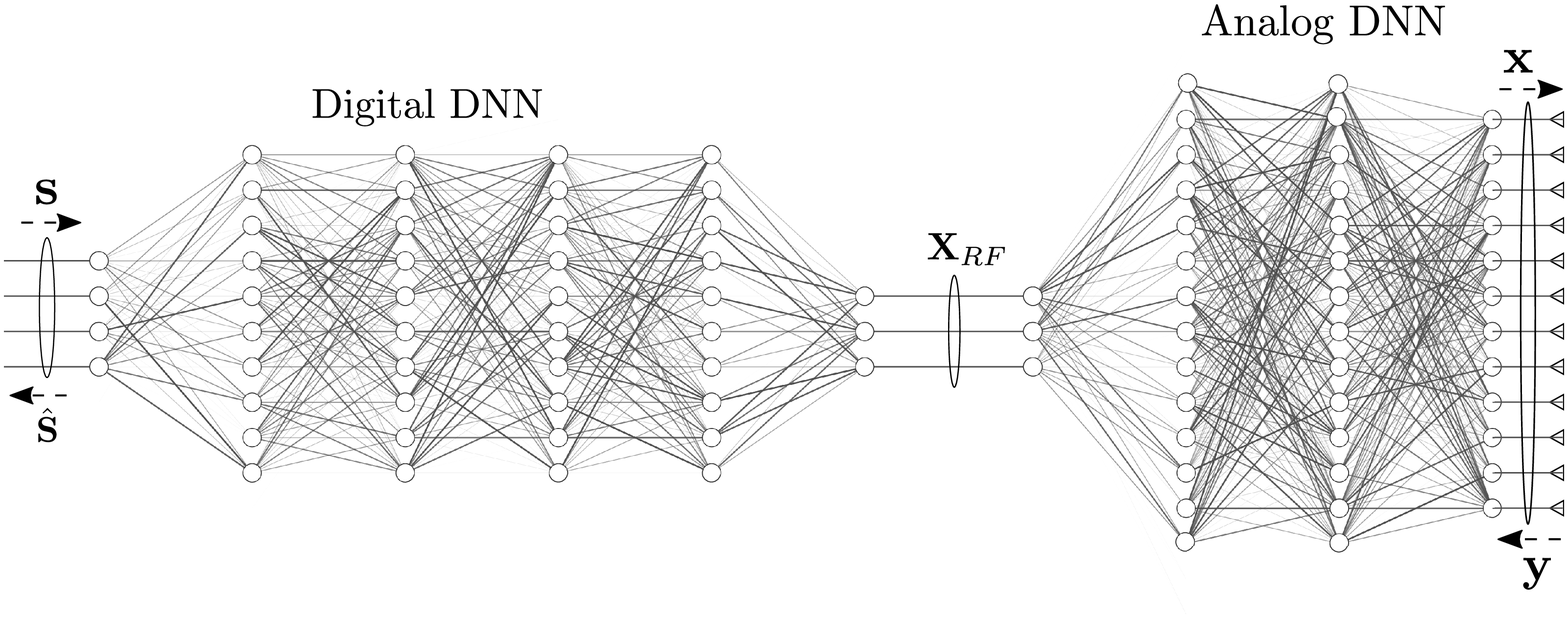}
\vspace{-4em}
\caption{HDNN-based \mm~transceiver.}
\label{fig:nn}
\vspace{-4em}
\end{figure}

To simplify the training process, we can consider the ADNN as additional layers to the digital DNN, which brings us to
\vspace{-1em}
\begin{equation}
  \label{eq:fuldnn1}
\xt=\dnn_T^A\Big(\big(\dnn_T^D(\s;\tet_T^{D})\big);\tet_T^A\Big).
\vspace{-1em}
\end{equation}
From \eqref{eq:dnn} and \eqref{eq:per1}, we can write the above equation as
\vspace{-1em}
\begin{equation}
  \label{eq:xtdnn}
\xt=\dnn_T^{H}(\s;\tet_{T}),
\vspace{-1em}
\end{equation}
where $\tet_T=\{\tet_T^A,\tet_T^D\}$ is the composite parameter set of the resulting downlink HDNN..

\emph{Uplink:}
In this case, the non-linear digital processor in \eqref{fdhpr1} corresponds to a first ADNN, whose output is the RF signal vector $\xrf$, i.e.,
\vspace{-1em}
\begin{equation}
  \label{eq:ydnn}
\far(\y)=\dnn^A_R(\y;\tet^A_{R}),
\vspace{-1em}
\end{equation}
where $\tet_T^{A}$ is the parameter set of the ADNN. % \deleted{ and $\zrf$ is the input of
The RF output of this ADNN is then converted to the baseband where it is represented by $\xbb$.
Subsequently, the non-linear operator in \eqref{fdhpr} is realized by means of a digital DNN to produce the desired output vector $\hat{\mathbf{s}}$, i.e.,
\begin{equation}
  \label{eq:yxdnn}
\fdr(\xbb)=\dnn^D_R(\xbb;\tet^D_{R}),
\end{equation}
where $\tet_T^{A}$ is the parameter set of the DNN.
Hence, proceeding as in the downlink case, we can write
\vspace{-1em}
\begin{equation}
  \label{eq:xrdnn}
\hat{\mathbf{s}}=\dnn_R^{H}(\y;\tet_{R})=\dnn_R^A\Big(\big(\dnn_R^D(\y;\tet_R^{D})\big);\tet_R^A\Big),
\vspace{-1em}
\end{equation}
where $\tet_R$ is the parameter set of the resulting uplink HDNN.

The proposed HDNN architecture is depicted in Fig.~\ref{fig:nn} where the forward propagation is shown for both downlink and uplink scenarios.
Here we considered regular DNNs, however, different network structures, and DL techniques such as CNNs, RNNs, etc., can be indeed used in the proposed framework. Since it is desirable to employ as few RF chains as possible, depending on the specific task the HDNN may suffer from a bottle necking effect in the flow of information from the digital DNN to the ADNN and vice versa.
In theory, this problem can be partly mitigated by increasing the width and depth of the ADNN or the digital DNN, as it has been shown that deep and wide neural networks are universal approximators \cite{hornik_91,park_91,lensho_93,huang_06,Mhaskar16,hanin2019,hanin2017}.

Conventional FD beamforming schemes are linear operations which are continuous functions and thus can be generated by the proposed structure. Moreover, although the universality of DNNs is only shown for continuous functions, the so-called \textit{expressive power} of the DNNs \cite{hornik_91,park_91,lensho_93,huang_06,Mhaskar16,hanin2019,hanin2017} enables them to successfully approximate a variety of functions which to our benefit, can be leveraged to realize existing communication techniques within the hybrid DNN-based systems.

\vspace{-1em}
\subsection{DL-based Transceiver Design}
\label{sec:FR}
In the following, we briefly discuss how the proposed framework can be used for HSP-based \mm~transceiver design. For the sake of brevity, we only discuss the downlink problem formulation but the uplink problem can be also formulated in a similar way.

The HDNN can be trained to learn an arbitrary transmission scheme, as defined by the mappings $\fdt(\cdot)$ and $\fat(\cdot)$ in \eqref{fdhp} and \eqref{fdhp1}. This can be done by training the HDNN using the following loss function,
\begin{equation}
  \label{eq:supervisedloss}
  L(\tet_{T})=\norm{\dnn_T^{H}(\s;\tet_{T})-\fat\circ\fdt(\s)}_2.
\end{equation}
This approach can be applied in principal to a variety of non-linear beamforming technique such as vector perturbation \cite{Hochwald_TCOM05}, Tomlinson-Harashima precoding \cite{Tomlinson1971,HarashimaTCOM72}, robust beamformer design under imperfect CSI \cite{Morsali_WCNC_19}, and space-time coding \cite{MorsaliWCOML19}.
In the next section, we design and train HDNNs with $\fdt(\s)$ selected as the eigen-mode beamforming for both uplink and downlink.

When the HDNN is trained using \eqref{eq:supervisedloss}, the output of the HDNN ideally appropriates the selected transmission scheme, i.e., $\fdt(\s)$.
However, to take a step further and exploit the power of machine learning paradigm, it is possible to let the transmitter learn the optimal transmission scheme without explicitly specifying its structure \emph{a priori}. To be more specific, the loss function for training the HDNN can be a performance measure of the overall communication link, i.e., taking the transmitter and receiver into account, as well as other constraints and system requirements. Depending on the setup, constraint and requirements of the system it is possible to formulate various loss functions.
For instance, in the case of a massive-MIMO BS with a single UE, and assuming that the latter UE uses a fixed combiner matrix $\mathbf{C}$ known at the transmitter side, the HDNN can be trained using,

\vspace{-2em}
\begin{equation}
  \label{eq:unsupervisedloss1}
 L(\tet_{T})=\norm{\mathbf{C}\,\Hd\,\dnn_T^{H}(\s;\tet_{T})-\s}_2.
\vspace{-1em}
\end{equation}
For a multi-user scenario, the signal-to-interference-plus-noise (SINR) can be used instead of the symbol error used above.
Furthermore, additional constraints can be considered such as imperfect CSI with norm-bounded error to design robust beamformers.

For sake of completeness, we also mention other machine learning techniques that can be used within the proposed HDNN framework.
Since in mobile and cellular communication systems, CSI changes rapidly, efficiency and computational complexity are important issues. Consequently, transfer learning techniques can be leveraged to facilitate the training process \cite{YangTCOM202}. Moreover, since cloud-ran is highly anticipated in B5G and 6G systems, distributed learning techniques (e.g. federated learning and split learning) can be used in designing and training of HDNNs \cite{Federated}.
Ultimately, the proposed framework could be used in a reinforcement learning (RL) setup to train an end-to-end communication system where channel estimation, beamforming, modulation, etc., are all performed naively and directly via RL.

\vspace{-1em}
\section{Supervised Learning-based HBF with FD Performance}
\label{Sec:HBF}
In this section, to illustrate the potential advantages of the proposed DL framework for HSP, we apply it to design supervised learning-based hybrid beamformers that achieve the performance of FD beamforming with limited number of RF chains. We focus on the single user although extension to multi-user is possible.

\vspace{-1em}
\subsection{Uplink and Downlink Beamforming with $\nrf\geq\ns$}
\label{subsec:Normal}
As discussed above, using the proposed framework, it is possible to closely approximate any mapping form symbol vector $\s$ to the desired transmitted signal $\xt$. Consequently, we can use supervised learning to train the DNN such that the hybrid system generates the output of the FD beamforming with limited RF chains.

In the downlink, the optimal eigen-mode precoding is obtained by solving the following problem:
\vspace{-2em}
\begin{subequations}\label{eq:FDoptprecoder}
	\begin{align}
	&\max_{\P}\quad\log_2 \det(\mathbf{I}_\nr+\Hd\P\P^H\Hd^H),\\
	&~\text{s.t.}\quad\quad\text{Tr}(\P\P^H)\leq 1.
\vspace{-1em}
	\end{align}
\end{subequations}
The solution to the above problem is given by $\P={\mathbf{V}}_D\mathbf{\Upsilon}_D$,
where the diagonal weight matrix $\mathbf{\Upsilon}_D$ is calculated via water filling \cite{TorkildsonITWC11} and ${\mathbf{V}}_D$ is a unitary matrix obtained from the singular value decomposition of the channel matrix, represented in standard form as,
$\Hd=\mathbf{U_D}\mathbf{\Sigma_D} {\mathbf{V}_D}^H$.
Thus, the desired output of the HDNN for symbol vector $\s$ is given by $\P\s$ and using \eqref{eq:xtdnn}, we wish to find $\tet_T$ such that $\dnn_T^{H}(\s;\tet_{T})=\P\s$.
We use an HDNN where the digital DNN has $L_D=5$ hidden layers of size $\ns\cba$ and the ADNN has $L_A=1$ hidden layer of size $\nt\cra$. These hyperparameters (which control the depth of the network and width of the hidden layers) may vary based on the number of antennas and the number of RF chains and can be obtained by cross referencing.

For each channel instance $\Hd$, the network is trained by generating random symbol vectors and their corresponding desired transmitted signal. The network is then trained for the dataset $\pazocal{D}_{\HH}=\{(\s^{(i)},\P\s^{(i)})\}_{i=1}^{N_D}$.
Since the DNNs must is trained for regression, selecting the right loss function is crucial, particularly, because in mmWave beamforming, a slight deviation can cause huge performance degradation. From an optimization perspective, the $\ell 1$ norm results in a more sparse error vector which is more favourable for our purpose compared to $\ell 2$ norm. Consequently, we propose to the mean absolute error loss function which can be expressed as
\begin{equation}
  \label{eq:loss}
 L(\tet)=\frac{1}{N_D}\sum_{\s\in\pazocal{B}_j}|\dnn_T^{H}(\s;\tet_{T})-\P\s|,
\end{equation}
for the mini-batch $\pazocal{B}_j$. Finally, using an optimization method such as Adaptive moment estimation (Adam) the weights and biases of the network are updated during the backpropagation phase.

For uplink beamfomer (combiner), a similar network structure with the same hyperparameters as the downlink case can be used as shown in Fig.~\ref{fig:nn}. The optimal FD combiner can be obtained from
\vspace{-1em}
\begin{equation}\label{eq:recRate}
\max_{\mathbf{C}}~\log_2\det\big(\mathbf{I}_\ns+{\rho}(\mathbf{C}\mathbf{C}^H)^{-1} \mathbf{C}\Hu\Hu^H\mathbf{C}^H\big).
\vspace{-1em}
\end{equation}
By writing the singular value decomposition of the uplink channel as $\Hu=\mathbf{U_U}\mathbf{\Sigma_U} {\mathbf{V}_U}^H $, we have,
$\mathbf{C}^H=\mathbf{U}_U^a$,
where $\mathbf{U}_U=[\mathbf{U}_U^a,\mathbf{U_U}^b]$ and $\mathbf{U_U}^a$ contains the first $\ns$ columns of $\mathbf{U}_U$, corresponding to the $\ns$ dominant singular values.
Training the DNN for uplink is not as straightforward as for the downlink because the received signal is dominated by noise. Consequently, we let
\vspace{-1em}
\begin{equation}
  \label{eq:uplinkinput}
 \y=\rho\z+\nvec,
\vspace{-1em}
\end{equation}
where $\z\sim\mathscr{CN}(\mathbf{0},\mathbf{I}_\nr)$ is a Gaussian random vector and we want to find $\tet_R$ such that $\dnn_R^{H}(\y;\tet_{R})=\mathbf{C}\y$, i.e., the network is trained for the dataset $\pazocal{D}_{\HH}=\{(\y^{(i)},\mathbf{C}\y^{(i)})\}_{i=1}^{N_D}$.
The same loss function and optimizer as in the downlink can be used for learning the network parameter.

\vspace{-1em}
\subsection{Downlink Beamforming with $\nrf<\ns$}
\vspace{-1em}
\label{sec:lsrf}
To illustrate the potential of the proposed framework, in this subsection we present a hybrid precoder which can achieve the same performance as the FD beamforming with less than $\ns$ RF chains.
From \eqref{eq:hpconv}, we can see that to achieve the same performance as the FD beamforming in conventional \hp, i.e., $\pfd=\pa\pd$, the minimum number of required RF chains must be equal to $\textrm{rank}(\pfd)$, where $\pfd\in\mathbb{C}^{\nt\times \ns}$. In practice, we usually have $\textrm{rank}(\pfd)=\ns$, which is why in the hybrid beamforming literature it is assumed that $\nrf\geq\ns$. Even with this assumption achieving the same performance as the FD beamforming is generally not possible \cite{AyachITWC14,AlkhateebJWCOM16,YuJSTSP,Mai_WCNC_16,Nguyen17,Morsali_WCNC_19,LiICL16,JiangWCOM0}, except under certain conditions \cite{MorsaliCOML17,MorsaliGSIP19,MorsaliICASSP20}.

Using the proposed framework, we design an HDNN where the digital DNN with $L_D=4$ layers of size $\ns\cbb$ is connected via $\ns/2$ RF chains to an ADNN with $L_A=4$ hidden layer of size $\nt\crb$ where $\cbb$ and $\crb$ are network parameters that control the width of hidden layers in baseband and analog neural networks, respectively.

The same training procedure as discussed in previous subsection is used to update the network parameters.
The proposed HBF with $\ns/2$ RF chains can achieve the same performance as a FD system as will be shown in the next section.

\begin{figure}[t]
	%	\vspace{-1em}
	\centering
	\includegraphics[width=0.7\linewidth]{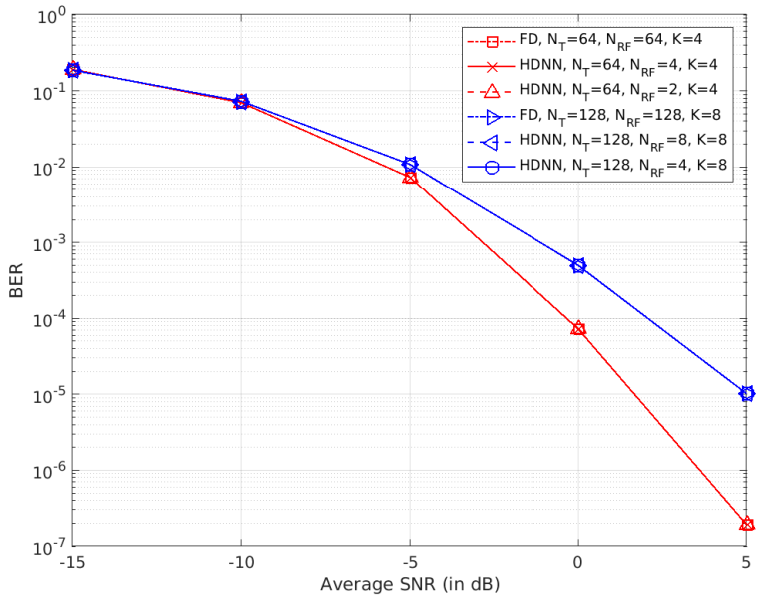}
	\vspace{-1em}
	\caption{BER versus SNR of proposed HDNN designs and FD beamforming for downlink connection of $64\times 4$ and $128\times 8$ systems.}
	\label{fig:figBerNT}
	\vspace{-6em}
\end{figure}

It is possible to design a similar HDNN for the uplink beamforming, however, in order to achieve the same performance as the FD combiner, due to pretense of noise, a more intricate DNN architecture and training process is required which remains a topic for future studies.

\begin{figure}[t]
	\centering
	\includegraphics[width=0.7\linewidth]{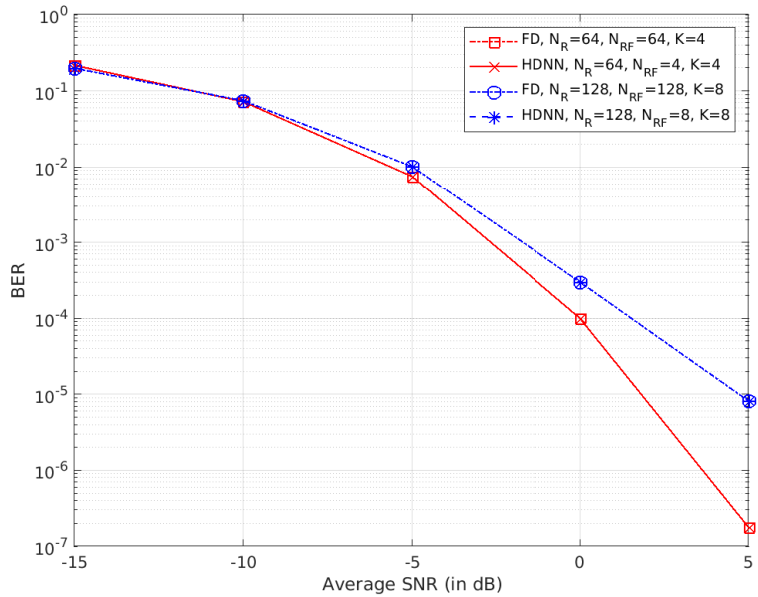}
		\vspace{-1em}
	\caption{BER versus SNR of proposed HDNN designs and FD beamforming for uplink connection of $64\times 4$ and $128\times 8$ systems.}
	\label{fig:figBerNr}
		\vspace{-3em}
\end{figure}

% \vspace{-4em}
\section{Simulation Results}
\label{sec:Simul}
In this section, simulation results are presented to illustrate the merits of the proposed HDNN in section~\ref{Sec:HBF}, comparison with FD beamforming, as well as benchmark hybrid designs from the literature are included.

\vspace{-2em}
\subsection{Methodology}
\vspace{-1em}

We use the following mmWave massive-MIMO channel model \cite{SohrabiISTSP16,LinJCOM19},
\begin{equation}
\mathbf {H} = \sqrt {\frac {{{\nt}{\nrue}}}{{{N_{\mathrm {c}}}{N_{\mathrm {ray}}}}}} \sum \limits _{i = 1}^{{N_{\mathrm {c}}}} {\sum \limits _{j = 1}^{{N_{\mathrm {ray}}}} {{\alpha _{ij}}} } {\mathbf {a}_{\mathrm {r}}}(\theta _{ij}^{\mathrm {r}}){\mathbf {a}_{\mathrm {t}}}{(\theta _{ij}^{\mathrm {t}})}^{H},
\end{equation}
where $N_{\mathrm {c}}=5$ is the number of clusters, and $N_{\mathrm {ray}}=10$ is the number of rays in each cluster.  Similar to \cite{SohrabiISTSP16,LinJCOM19}, the path gains are independently generated as $\alpha_{ij}\sim\mathscr{CN}(0,1)$. The transmit and receive antenna responses are denoted by $ {\mathbf {a}_{\mathrm {r}}}(\theta _{ij}^{\mathrm {r}})$ and ${\mathbf {a}_{\mathrm {t}}}{(\theta _{ij}^{\mathrm {t}})}$ respectively. For simplicity, a uniform linear array of size $N$ with half-wavelength spacing is employed, hence  where,
\begin{equation}\label{Arrayres}
\mathbf{a}(\phi)=\frac{1}{\sqrt{N}}[1,e^{j\pi\sin(\phi)},\ldots,e^{j(N-1)\pi\sin(\phi)}].
\end{equation}
The angles of arrival $\theta _{ij}^{\mathrm {r}}$ and departure $\theta _{ij}^{\mathrm {t}}$ are independently generated according to the Laplacian distribution with mean cluster angles $\bar{\theta}_{ij}^{\mathrm {r}}$ and $\bar{\theta} _{ij}^{\mathrm {t}}$, uniformly distributed in $[0,2\pi]$, while the angular spread is set to 10 degrees within each cluster. We further assume that the channel estimation and system synchronization are perfect.

\begin{figure}[t]
%	\vspace{-1em}
	\centering
	\includegraphics[width=0.7\linewidth]{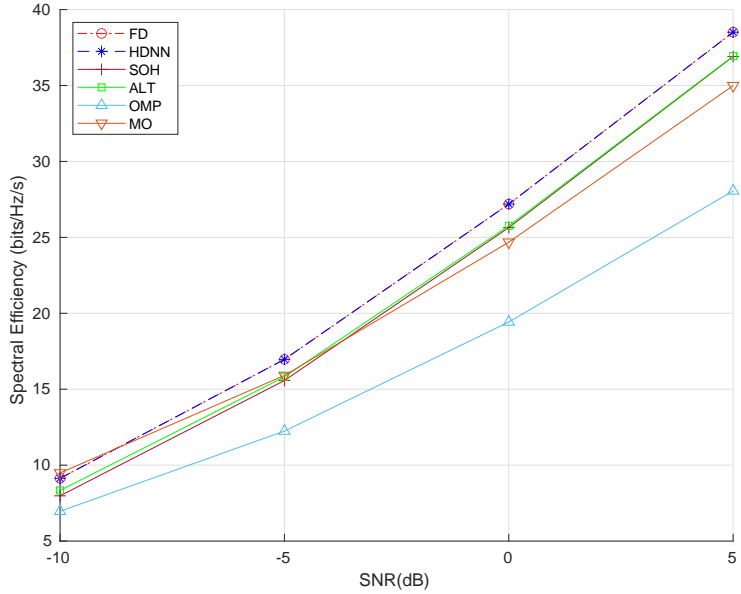}
		\vspace{-1em}
	\caption{Spectral efficiency versus SNR of different methods for downlink connection of $\nt=128$ \mm~BS and a UE with $\nrue=8$ antennas.}
	\label{fig:don128}
		\vspace{-5em}
\end{figure}

The parameters of the proposed HDNN are set as follows: For the case $\nrf=\ns$, the layer configuration parameters are set to $L_D=5$, $L_A=1$ and $\cba=2$, $\cra=3$. For the case $\nrf=\ns/2$, the model parameters are set to $L_D=4$, $L_A=4$, $\cbb=2$ and $\crb=6$.
The HDNN in the downlink is trained with 500000 random symbols vectors as the inputs and their corresponding precoded signal vectors as the desired outputs for each channel realization.
For uplink training, due to existence of noise, using \eqref{eq:uplinkinput}, 500000 noisy gaussian random vectors are considered as the inputs and the desired outputs are the the corresponding combined vectors, i.e. $\mathbf{C}\y$.
For each channel realization, training is performed by the Adam optimizer with learning rate $0.001$ for 5 epochs with mini-batch size of 50.

\subsection{Results and Discussion}

We present the spectral efficiency (SE) and bit error rate (BER) performance of the proposed HDNN design and the FD beamforming to confirm that the former achieves same performance as the latter. However, since our design achieves the same performance as the FD beamforming, one can find comparison between our design and existing hybrid designs directly in the corresponding work, e.g., \cite{SohrabiISTSP16,YuJSTSP,Nguyen17,LinJCOM19,AlkhateebAccess18, HuangTVT19,  LinWCOML20, SungACESS20}.
\begin{figure}[t]
	\centering
	\includegraphics[width=0.7\linewidth]{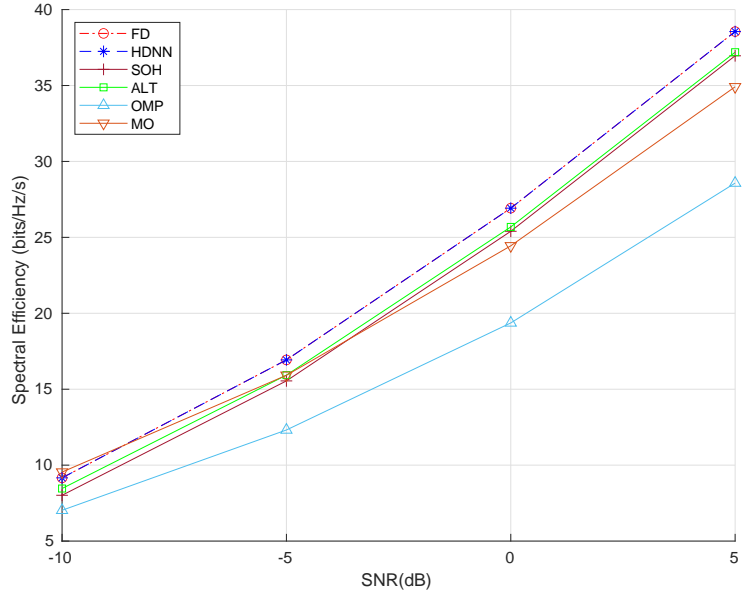}
		\vspace{-1em}
	\caption{Spectral efficiency versus SNR of different methods for uplink connection a UE with $\nrue=8$ and a \mm~BS with $\nr=128$.}
	\label{fig:up128}
		\vspace{-5em}
\end{figure}

The BER performance versus signal-to-noise ratio (SNR) (which corresponds to parameter $\rho$ in \eqref{eq:hpconv} under normalization of the additive noise) for $64\times 4$ and $128 \times 8$ point-to-point MIMO setups are illustrated for uplink and downlink connections in Fig.~\ref{fig:figBerNT} and~\ref{fig:figBerNr}, respectively. In both cases the multi-antenna UE performs FD beamforming, 4-QAM constellation is used and the number of transmitted symbols is equal to the number of UE's antennas, i.e., $\ns=\nrue$.
It can be observed that the proposed HDNN design matches the performance of the FD system with only $\ns$ RF chains. Moreover, for downlink beamforming, our proposed design in Subsection~\ref{sec:lsrf} achieves the same performance with $\nrf=\ns/2$ RF chains which to the best of our knowledge has not been reported in the literature before.
\begin{figure}[t]
%	\vspace{-1em}
	\centering
	\includegraphics[width=0.7\linewidth]{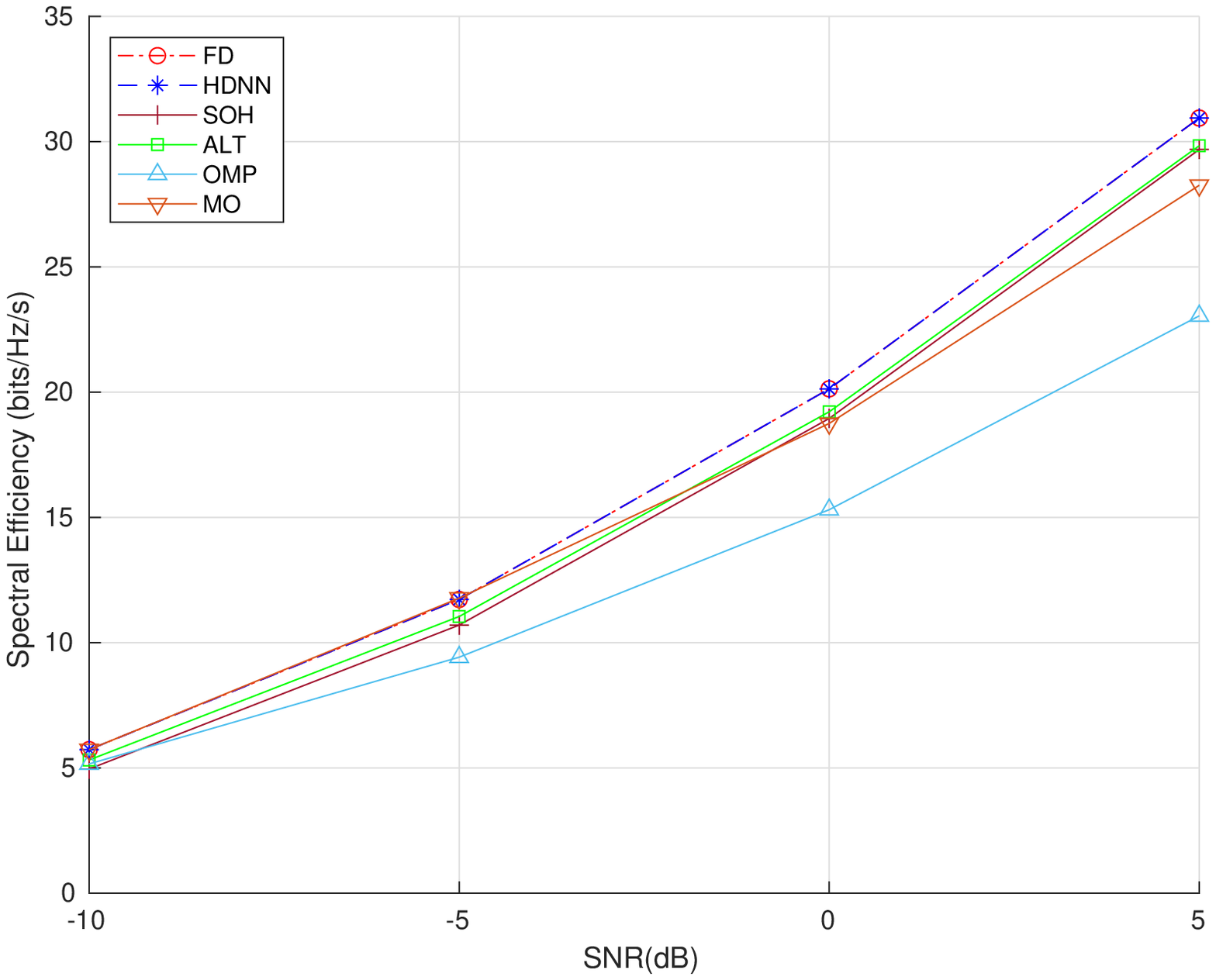}
	\caption{Spectral efficiency versus SNR of different methods for downlink connection of $\nt=64$ \mm~BS and a UE with $\nrue=4$ antennas.}
	\label{fig:don64}
		\vspace{-5em}
\end{figure}

Next, we compare the SE of our proposed HDNN design in Subsection~\ref{subsec:Normal} to FD beamforming as well as benchmark HBF designs in, \cite{LinJCOM19}, \cite{SohrabiISTSP16}, \cite{YuJSTSP}, and \cite{Nguyen17}, which are referred to as MO, SOH, ALT, and OMP, respectively.
The SE of the downlink and uplink connections when a \mm~BS with $\nt=128$ serves a UE with $\nrue=8$ antennas for $\ns=8$ are shown in Fig.~\ref{fig:don128} and Fig.~\ref{fig:up128}, respectively. While the FD system requires $\nrf=128$ RF chains, hybrid beamforming techniques only need $\nrf=8$ RF chains. As shown in the figures, for MO, SOH, ALT, and OMP HBF the cost of having less RF chains is degraded performance compared to FD systems, whereas, the proposed HDNN matches the rate of the FD beamforming with the same number of RF chains as the existing hybrid designs.

\begin{figure}[t]
%	\vspace{-1em}
	\centering
	\includegraphics[width=0.7\linewidth]{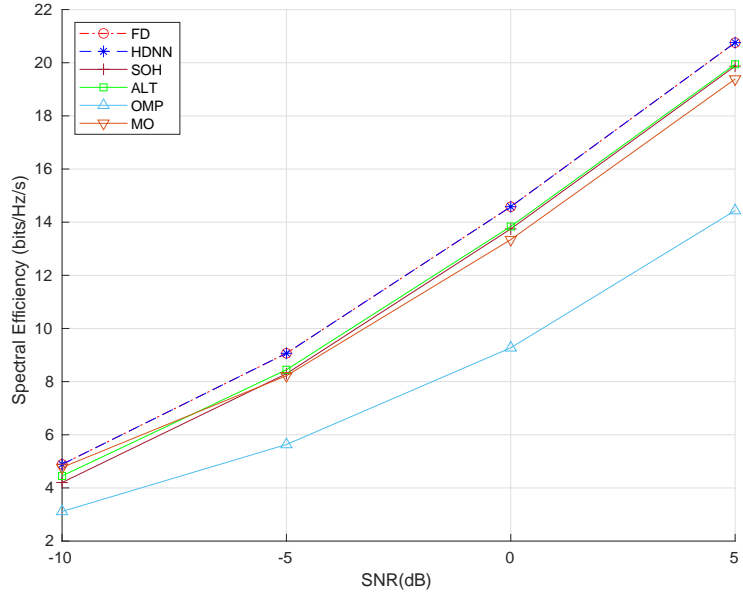}
	\vspace{-1em}
	\caption{Spectral efficiency versus SNR of different methods for uplink connection a UE with $\nrue=4$ and a \mm~BS with $\nr=64$.}
	\label{fig:up64}
		\vspace{-2em}
\end{figure}

Moreover, downlink SE of a \mm~BS with $\nt=64$ serving a UE with $\nrue=4$ antennas for $\ns=4$ is illustrated in Fig.~\ref{fig:don64}. It can be observed that the proposed HDNN matches the SE of the FD beamforming while outperforming the benchmark HBF designs. The SE of the uplink connection for the same system configuration used in reverse direction (uplink) is also depicted in Fig.~\ref{fig:up64}. As in the downlink case, the proposed HDNN design achieves the same rate as the FD systems and has higher rate than the existing hybrid designs.

\section{Conclusion}
In this paper, we presented a general DL framework for efficient design and implementation of \hp~in \mm~systems. By exploiting the fact that any complex matrix can be written as a scaled sum of two matrices with unit-modulus entries, we first presented a novel ADNN which can be implemented with common RF.
This structure was subsequently embedded into an extended HDNN architecture to facilitate the implementation of \mmw~\mm~systems and improving their performance. Since the proposed HDNN architecture enables \hp-based \mm~transceivers to approximate any desired transmitter and receiver mapping with arbitrary precision, we were able to present a new HDNN-based beamformer design that can achieve the same performance as FD beamforming, with reduced number of RF chains.
Simulation results were finally presented which confirmed that our design outperforms benchmark hybrid design in the literature by achieving the same performance as the FD systems.

\renewcommand{\baselinestretch}{1}
\bibliographystyle{IEEEtran}
\bibliography{IEEEabrv,Biblo}

\end{document}